\documentclass[twocolumn, amssymb, amsmath,nobibnotes,nofootinbib, aps, prb,showpacs]{revtex4-2}
\usepackage[dvips]{graphicx}
\usepackage{amsmath}
\usepackage{bm}
\usepackage{booktabs}
\usepackage{color}
\begin{document}
	\title{Magnetic Metastability driven Anomalous Hall Effect in Fe$_{x}$TaS$_2$ }
	\author{Mohamad Numan$^1$, Prasanta Chowdhury$^1$, Sanat Kumar Adhikary$^2$, Saurav Giri$^1$, Jhuma Sannigrahi$^3$, Matthias Gutmann$^4$, Souvik Chatterjee$^2$, and Subham Majumdar$^1$}
	\email{sspsm2@iacs.res.in}
	\affiliation{$^1$School of Physical Science, Indian Association for the Cultivation of Science, 2A \& B Raja S. C. Mullick Road, Jadavpur, Kolkata 700 032, India}
	\affiliation{$^2$UGC-DAE Consortium for Scientific Research, Kolkata Centre, Sector III, LB-8, Salt Lake, Kolkata 700 106, India}

        \affiliation{$^3$School of Physical Sciences, Indian Institute of Technology Goa, Farmagudi, Goa 403401, India}

        \affiliation{$^4$ISIS Neutron and Muon Source, Science and Technology Facilities Council, Rutherford Appleton Laboratory, Chilton Didcot OX11 0QX, United Kingdom}

\begin{abstract}
	We report here the observation of robust thermoremanent magnetization (TRM) in Fe-intercalated TaS$_2$ single crystal Fe$_{x}$TaS$_2$, which develops upon cooling under a moderate magnetic field of 500 Oe or higher. The thermoremanence is found to exist in a zero magnetic field up to a temperature as high as 160 K, which is well above the ferromagnetic Curie point ($T_C =$ 90 K) of the sample. The TRM in this compound can be traced back to the strong Ising character and the quasi-two-dimensional nature of the spins. The analysis of the magnetization data indicates the presence of short-range magnetic correlation in the form of a Griffiths singularity above $T_C$ to a temperature as high as 2$T_C$. The TRM is further manifested in the form of a large anomalous Hall effect (AHE) in zero magnetic field. In contrast to TRM, which exists well above $T_C$, the thermoremanent Hall effect vanishes as soon as the sample is heated above $T_C$. This may be an indication that a long-range ordered phase is essential for the emergence of AHE. 	
\end{abstract}	
\maketitle
	
	
\section{Introduction}
Layered van der Waals (vdW) materials continue to garner significant attention due to their fascinating physical properties, both from a fundamental physics perspective and for their potential applications in various two-dimensional (2D) spintronics devices~\cite{Stojchevska_science,Lin2019,Manzeli2017}. Among them, transition metal dichalcogenides with general formula TX$_2$ (T = 4$d$ or 5$d$ transition metal, and X = chalcogen) are of particular interest on account of their novel electronic, optical, and mechanical properties. In terms of electronic properties, TX$_2$ exhibits various many-body phenomena, including superconductivity~\cite{Sipos_nmat}, charge density wave (CDW) instabilities~\cite{Wilson_CDW,Thomson_CDW,Yoshida_CDW}, and even a quantum spin liquid state~\cite{Lee_PNAS,Klanjšek_nphys,Manas_npjQM}. 

\par
The quasi 2D structure of TX$_2$ enables us to intercalate various metal atoms in the vdW gap between two TX$_2$ layers. Very recently, 3$d$ transition metal intercalated TX$_2$ compounds emerged as a test bed for studying topological effects related to magnetism, which include chiral magnetic soliton lattice arising from helimagnetic structure~\cite{Togawa2012}  and a giant planar Hall effect in Cr$_{1/3}$NbS$_2$~\cite{Mayoh2022}, as well as a spontaneous topological Hall effect induced by non-coplanar antiferromagnetic order in  Co$_{1/3}$NbS$_2$ and Co$_{1/3}$TaS$_2$~\cite{Takagi2023}. The long-range order and structural properties of these intercalated materials strongly depend on the specific 3$d$ T-metal and the percentage of intercalation ($x$)~\cite{Eibshutz_1981,Dijkstra_1989,Narita_1994}.
\par 
Among various intercalated TX$_2$, Fe$_{x}$TaS$_2$ displays unique magnetic properties. For intercalation levels $0.23 \leq x \leq 0.4$, the material exhibits ferromagnetic (FM) order. In particular, the Curie point ($T_C$) increases as $x$ decreases, although no perfect correlation has been identified, as $T_C$ continues to rise even with further reductions in $x$~\cite{Chen_2016}. At specific intercalation levels, such as $x = 1/4$ or $x = 1/3$, the intercalated Fe ions arrange themselves into a superlattice. This superlattice exhibits periodic arrangements, forming 2$a \times$ 2$a$ or $\sqrt{3} a \times$ $\sqrt{3} a$ superstructures, respectively. 

\par
In recent times, there has been a growing interest in understanding the anomalous Hall effect (AHE), and these 3$d$ metal intercalated materials provide an excellent platform for exploring their electronic states. Checkelsky {\it et al.} performed detailed Hall and magneto-transport studies on Fe$_{1/4}$TaS$_2$, and they observed large AHE in the system, where the intrinsic Berry phase mechanism dominates below 50 K~\cite{Checkelsky_2008}.
\par
Several 3$d$ transition-metal intercalated dichalcogenides including Fe$_x$TaS$_2$ show quasi two-dimensional spin structure with strong out-of-plane magnetic anisotropy. Despite numerous works, there are very few reports~\cite{Park2022} addressing metastability associated with the magnetic state. Here, we focus on the single-crystalline sample of Fe$_{0.27}$TaS$_2$  with particular attention to the effect of the field-cooled state and its evolution as a function of temperature. We find that cooling in a moderate magnetic field ($H$) of 500 Oe down to 3 K, the system achieves magnetization as high as its saturation value, and it holds it even when the field is removed. This metastable state persists up to a temperature as high as 2$T_C$, keeping the system magnetically trapped under weak field cooling. Most importantly, this metastable state with non-zero magnetization produces a robust AHE without any external field below $T_C$.

\begin{figure}
	\centering
	\includegraphics[width = 8.5 cm]{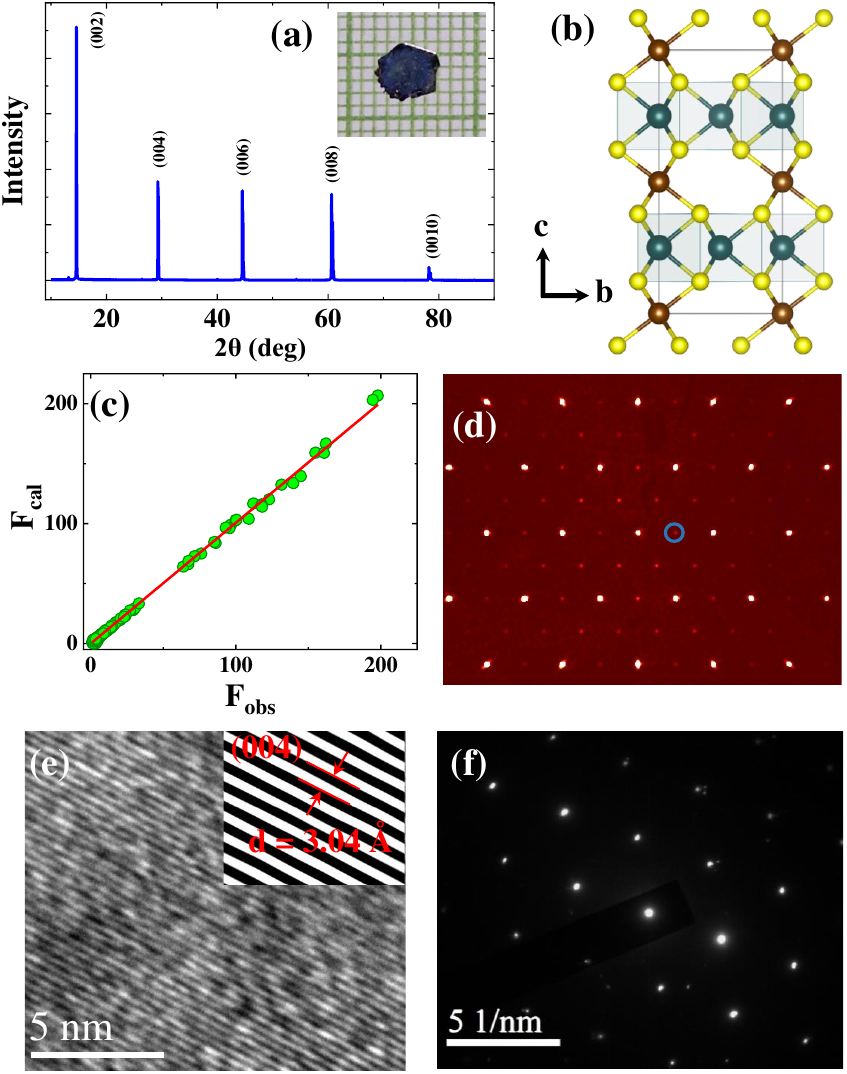}
	\caption{(a) XRD pattern in the flat surface of the crystal, showing sharp peaks corresponding to (0 0 2$l$) reflections, confirming its highly oriented layered structure. The inset displays an optical image of the sample crystal placed on to a 1 mm $\times$ 1 mm graph paper. (b) Crystal structure schematic of the material, highlighting the layered arrangement with the Fe atoms (brown colored) intercalated in between van der Waals gaps. (c) Observed versus calculated structure factors (F$_\text{obs}$ vs. F$_\text{cal}$), confirming the reliability of the structural refinement. (d) Single crystal x-ray diffraction pattern confirming the crystallinity of the sample with distinct diffraction spots in the $ab$ plane of the reciprocal space. (e) HRTEM image showing lattice fringes with an interplanar spacing of 3.04 \AA, as shown in the inset. (f) Selected area electron diffraction (SAED) pattern, with a marked spot corresponding to a specific reflection.}
	\label{xrd}
\end{figure}

\section{Experimental Methods}
Single crystals of Fe${_x}$TaS${_2}$ were synthesized using the standard chemical vapor transport method, with iodine as transport agent. The resulting crystals exhibit a flat plate-shaped geometry along the $ab$ plane, as shown in the inset of Fig.~\ref{xrd} (a), with typical dimensions of several millimeters. The crystal structure and phase purity were verified by x-ray diffraction (XRD) in the Bragg-Brentano $\theta$-$\theta$ geometry and the single-crystal x-ray diffraction (SCXRD). XRD on the flat plane $ab$ was performed using a RIGAKU SmartLab system (4.5 kW) with Cu $K_{\alpha}$ radiation (wavelength $\lambda$ = 1.5406 \AA). SCXRD data were collected using a Rigaku-Oxford Diffraction Xtalab Synergy single-crystal diffractometer equipped with a HyPix hybrid pixel array detector and Mo $K_{\alpha}$ radiation. Data reduction, structure solution, and refinement were performed using CrysAlisPRO, SHELXT~\cite{Sheldrick}, and JANA2020~\cite{JANA} software respectively.
\par
The stoichiometry of the compound was confirmed by energy-dispersive x-ray (EDX) analysis using a JEOL JSM-6010LA scanning electron microscope (SEM). High-resolution transmission electron microscopy (HRTEM) and selected area electron diffraction (SAED) were conducted with a JEOL TEM 2010. Magnetic measurements were performed using a Quantum Design Magnetic Properties Measurement System (MPMS3). Electrical transport measurements ($\rho_{xx}$ and $\rho_{yx}$) were performed with a cryogen-free high magnetic field system from Cryogenic Ltd., UK.

\begin{table}[h!]
\centering
\caption{Crystallographic and Structural Parameters for Fe$_{x}$TaS${_2}$}
\setlength{\tabcolsep}{0.15cm}
\begin{tabular}{@{}lll@{}}
\toprule
\textbf{Category}             & \textbf{Parameter}                      & \textbf{Value}                         \\ \midrule
                              & Crystal System                          &  Hexagonal                             \\
                              & Space Group                             &  \textit{P}6$_3$/mmc                   \\
\textbf{Crystal}              & Unit Cell                               & a =  6.6243(2) \AA                     \\
\textbf{System}               & Dimensions                              & c =  12.1878(4) \AA                    \\
                              & Unit Cell Volume                        &  V = 463.17(2) \AA$^3$                 \\
                              & Z                                       & 8                                      \\
                              & Density (cal)                           & 7.4293 g/cm³                           \\ \midrule
                              & Radiation                               & Mo $K_\alpha$, $\lambda$ = 0.71073 \AA \\
                              & Temperature                             & 295 K                                  \\
\textbf{Experimental}         & 2$\Theta$ Range                         & 3.35$^\circ$ to 65.32$^\circ$     \\
\textbf{Details}              & Refl. Collected                         & 10303                               \\
                              & Ind. Reflections                        & 357                               \\
                              & G.O.F.                                  & 1.97                                \\
                              & R Index for                             & R$_1$ = 0.036                       \\
                              & all data                                & wR$_2$ = 0.069                      \\\midrule
\textbf{Atomic}     & Atom Positions & See below                            \\ 
\textbf{Parameters}    & (x, y, z) and U$_\text{ani}$ & See below                            \\ \midrule
\end{tabular}

\vspace{0.5cm}
\setlength{\tabcolsep}{0.2cm}
\begin{tabular}{@{}lllll@{}}
\toprule
\textbf{Atom} & \textbf{x}      & \textbf{y}      & \textbf{z}      & \textbf{U$_\text{ani}$ (\AA$^{2}$)} \\ \midrule
Ta1           &  0.4970(1)      &  0.5029(9)      & 3/4             & 0.006(9)                     \\
Ta2           & 0               & 0               & 0               & 0.005(3)                     \\
Fe            & 0               & 0               & 0               & 0.031(3)                     \\
S1            & 2/3             & 1/3             & 0.1200(6)       & 0.006(4)                     \\
S2            & 0.8325(3)       & 0.1674(9)      & 0.6219(8)        & 0.006(4)                     \\ \bottomrule
\end{tabular}
\label{tab:xrd}
\end{table}

\section{Results}

\subsection{Structural and Phase Analysis}
The SCXRD data at room temperature show that Fe$_{x}$TaS$_2$ forms a hexagonal crystal structure with the space group \textit{P}6$_3$/mmc. The linear correlation between the observed (F$_\text{obs}$) and calculated (F$_\text{cal}$) structure factors, shown in Fig.~\ref{xrd} (c), indicates satisfactory structural refinement. The refined crystallographic and statistical parameters are listed in Table~\ref{tab:xrd}. 

\par
The XRD pattern, obtained in the Bragg-Brentano $\theta$-$\theta$ geometry on the flat surface of the crystal [see Fig.~\ref{xrd} (a)], displays peaks corresponding to the (0 0 2$l$) planes, confirming that the flat surface represents the $ab$ plane. Fig.~\ref{xrd} (b) illustrates the arrangement of Fe atoms (brown), which are intercalated in the vdW gaps between TaS$_2$ layers, forming a triangular network in the $ab$ plane. Furthermore, the SCXRD pattern on ($h$ $k$ 2) layer [Fig.~\ref{xrd} (d)] reveals less intense spots, marked with blue circles, which correspond to 2$a \times$ 2$a$ superstructures previously reported in the literature~\cite{Morosan2007}.

\par
The SEM energy-dispersive spectra confirm a compositional ratio of $\text{Fe}:\text{Ta}:\text{S} = 0.27:1:1.98$, indicating an Fe content close to 0.25, consistent with the crystal structure derived from SCXRD. The HRTEM image in Fig.~\ref{xrd} (e) shows an interplanar spacing of 3.04~\AA\ for the (0 0 4) planes. The layered morphology and crystalline nature of Fe$_{x}$TaS$_2$ are further verified by SAED patterns, as shown in Fig.~\ref{xrd} (f). The lattice parameters obtained from SAED and XRD are in close agreement with each other. All measurements were performed on a unique piece of single-crystalline sample of the stoichiometry Fe$_{0.27}$TaS$_2$.

\begin{figure}
	\centering
	\includegraphics[width = 8.5 cm]{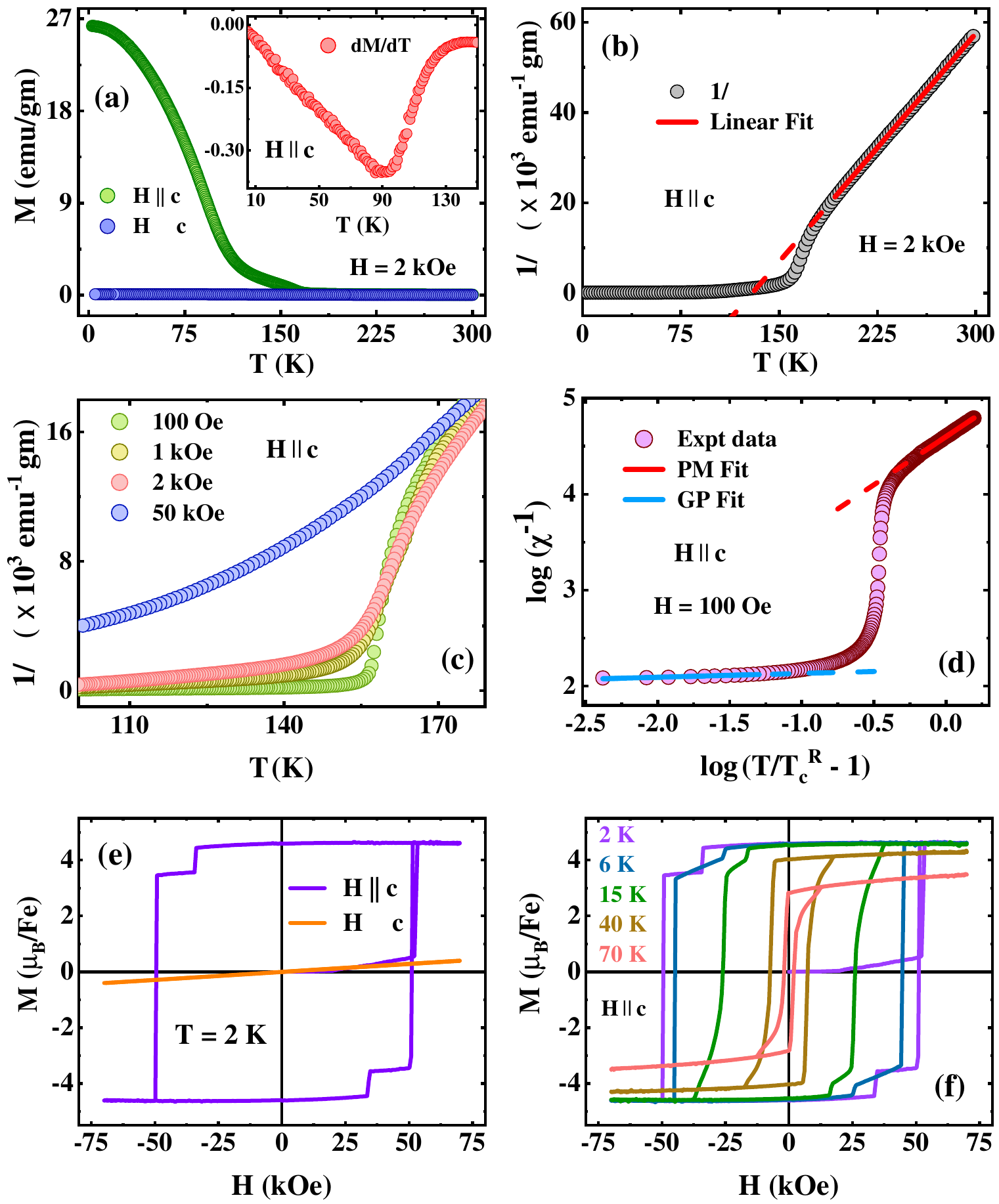}
	\caption{(a) Temperature-dependent magnetization ($M$-$T$) measured in field-cooled (FC) protocol under an applied field of 2 kOe with $H$ $\parallel$ $c$ and $H$ $\perp$ $c$ configurations. The inset shows the dM/dT plot with $T$ under $H =$ 2 kOe, when $H$ $\parallel$ $c$. (b) The inverse susceptibility (1/$\chi$) as a function of $T$. The red line displays linear fit to the data. (c) Temperature dependence of $\chi^{-1}$ under different applied fields. The plot highlights the deviation from the Curie-Weiss law at lower fields, which diminishes as the applied field increases. (d) Log-log plot of $\chi^{-1}$ as a function of ($T/T_{c}^R$-1). The solid lines represent linear fits in the paramagnetic (blue line) and Griffiths Phase (green line) regions, while the dashed extensions indicate the range over which the linear fits hold. (e) Isothermal magnetization ($M$-$H$) loop at 2 K for $H$ $\parallel$ $c$ and $H$ $\perp$ $c$ configurations, highlighting magnetic anisotropy along $c$ axis. (f) Temperature-dependent $M$-$H$ loops at selected temperatures (2 K, 6 K, 15 K, 40 K, and 70 K), showing the evolution of magnetic hysteresis with $T$ when $H \parallel c$.}
	\label{mag}
\end{figure}

\subsection{Magnetization}
\label{sec:magnetization}
Figure~\ref{mag} (a) presents the temperature ($T$) dependence of magnetization ($M$) under field-cooled (FC) protocol at an applied field of 2 kOe along the $c$ axis (green symbols) and perpendicular to the $c$ axis (blue symbols) for Fe$_{0.27}$TaS$_2$. As $T$ decreases from 300 K, $M$ initially increases slowly. Below approximately 160 K, $M$ increases rapidly, with a sharper increase observed below about 100 K for $H \parallel c$. In contrast, when $H \perp c$, the value of $M$ is significantly lower indicating a strong anisotropy in the system. The inset of Fig.~\ref{mag} (a) shows the first derivative of $M$ with respect to $T$, which exhibits a dip around 90 K, marking the long-range ordering temperature, $T_C$ of the system.

\par
Figure~\ref{mag}(b) presents a linear fit to the inverse susceptibility ($\chi^{-1}$, where $\chi = M/H$), confirming a FM ground state with a Curie-Weiss (CW) temperature of $\theta_{CW} = 129$ K. A significant deviation from linearity appears below 200 K, likely due to short-range correlations in the system. This deviation manifests itself as a downward shift from the expected CW behavior in the $\chi^{-1}$ vs $T$ plot.

\par
Notably, the downward deviation decreases with increasing $H$ and nearly vanishes at $H = 50$ kOe, shown in Fig.~\ref{mag} (c). The presence of such a deviation well above the magnetic ordering temperature ($T_C = 90$ K) is a key signature of a Griffiths-like phase (GP)~\cite{Pramanik2010}. The GP arises from the coexistence of finite-sized FM clusters within a paramagnetic (PM) background up to a characteristic temperature $T_G$ ($T_G > T_C$). The Griffiths temperature ($T_G$) is identified as the point where $\chi^{-1}(T)$ begins to deviate downward.

\par
The inverse susceptibility in the  GP obeys a power law behavior~\cite{Neto1998}: 
\begin{equation}
 \chi^{-1} \propto (T-T_{c}^R)^{(1-\lambda)}
\label{eqn:griffithsfit}
 \end{equation}
 Here, $\lambda$ is an exponent lying between 0 and 1, and and $T_{c}^R$ represents the critical temperature of a random FM clusters, at which susceptibility exhibits a tendency to diverge.

\par
To determine $\lambda$, a log-log plot of $\chi^{-1}$ vs ($T/T_{c}^R-$1) is used, from which the slope of the fitted straight line is extracted, as illustrated in Fig.~\ref{mag} (d). The slope within the GP region defines $\lambda_{\text{GP}}$, while that in the PM state determines $\lambda_{\text{PM}}$.

\par
An essential step in this analysis is the precise determination of $T_{c}^R$, because an incorrect value may result in unphysical fitting and inaccurate estimation of $\lambda$. To address this, the initial assumption is $T_{c}^R = T_C$, and $\lambda_{\text{PM}}$ is computed accordingly~\cite{Pramanik2010}. Subsequently, $T_{c}^R$ is fine-tuned in the fitting process until $\lambda_{\text{PM}}$ approaches zero, ensuring that the system transitions into a conventional PM phase above $T_G$, where it adheres to the CW fitting. Using this optimized $T_{c}^R$, we extract $\lambda_{\text{GP}}$, which quantifies deviations from the CW behavior due to order parameter non-analyticity. In this study, we determine the optimized values as $T_{c}^R = 117$ K and $\lambda_{\text{GP}} =$ 0.96. This value is close to 1 (maximum possible value), which establishes very strong Griffiths-like short range correlation~\cite{Salamon2002,Jiang2007,Magen2006}.

\begin{figure}
	\centering
	\includegraphics[width = 7 cm]{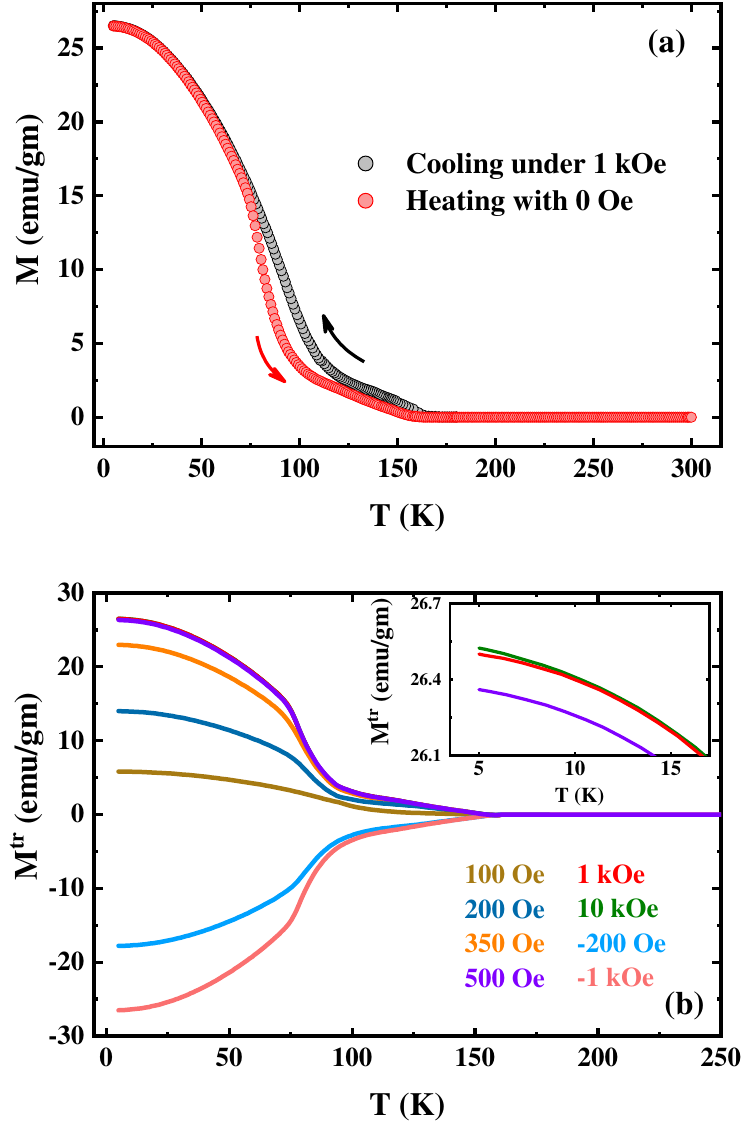}
	\caption{(a) $T$ dependent magnetization measured during cooling from 300 K under $H_{cool} =$ 500 Oe (black circles) and subsequent heating to 300 K in zero-field (red circles). The arrows indicate the direction of $T$ change. (b) $M$-$T$ curves measured during heating in zero-field after field cooling under various positive and negative cooling fields, demonstrating the thermoremanent magnetization. The inset highlights the low-temperature region, showing subtle variations in $M$ for $H_{cool}\geq$ 500 Oe.}
	\label{TRM}
\end{figure}

\par
The isothermal magnetization curves at 2 K, with $H$ applied along and perpendicular to the $c$ axis, are shown in Fig.~\ref{mag} (e).  For $H \parallel c$, $M$ initially increases linearly with $H$. However, above 50 kOe, $M$ increases sharply and subsequently saturates above 52 kOe with saturation moment $M_s=$ 4.6 $\mu_B$/Fe. Upon decreasing the magnetic field, a high remnant magnetization is observed and $M$ turns sharply negative when the field exceeds $-$49 kOe.  A large hysteresis loop is evident, with a coercivity of about 50 kOe. In contrast, for $H \perp c$, the $M$-$H$ curve remains linear and shows no sign of saturation even up to 80 kOe. In particular, the magnetization in this direction is significantly smaller compared to $H \parallel c$. This confirms the strong anisotropy in the material, with the $c$ axis being the easy axis and the $ab$ plane acting as the hard plane.

\par
Fig.~\ref{mag} (f) shows several $M$-$H$ isotherms recorded at different temperatures with $H \parallel c$. As $T$ increases, the coercivity of the sample decreases and it attains a value as low as 8 kOe at 70 K. The value of $M_s$ also decreases with increasing $T$. 

\begin{figure}
	\centering
	\includegraphics[width = 7.5 cm]{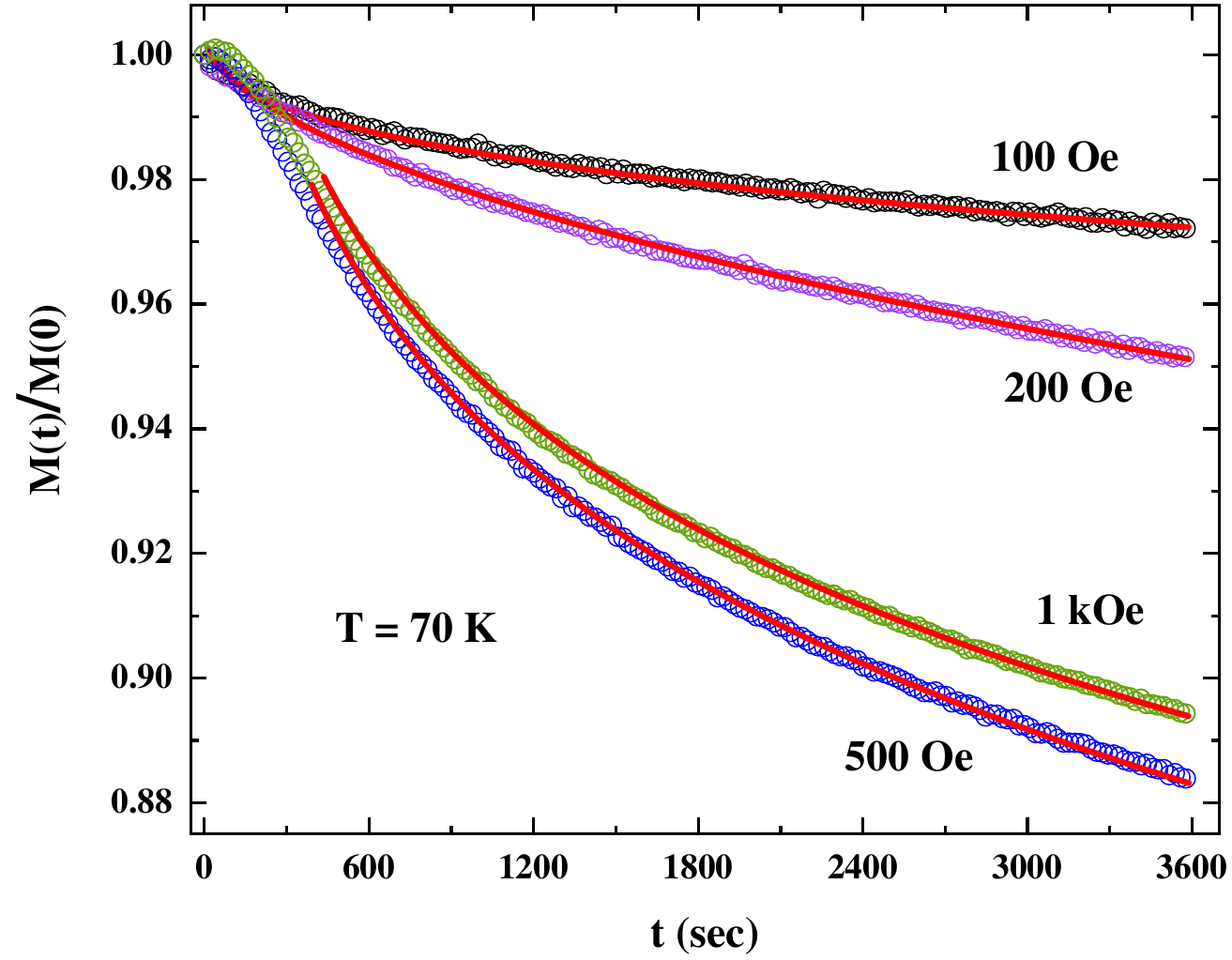}
	\caption{Normalized relaxation $M(t)/M(0)$ as a function of time at 70 K. The measurement was performed by cooling the sample to 70 K from 200 K followed by recording the relaxation over the course of an hour after the removal of applied field.}
	\label{relax}
\end{figure}

\par
The presence of a square $M$-$H$ loop and remanent magnetization ($M_R$) with large coercivity probably reflects the significant magnetic anisotropy and the Ising nature of the spins~\cite{Huang2017}. In particular, the value of $M_R$ remains substantial even at higher temperatures. This highlights the importance of investigating the evolution of remanent magnetization as a function of temperature, which is generally referred to as thermoremanent magnetization (TRM)~\cite{Chamberlin1984,Nordblad1986,mydosh1993,Mathieu2001}. The sample is cooled in an applied magnetic field below a characteristic temperature $T_0$, which may be the spin freezing temperature for the spin glass or the magnetic ordering temperature for the long-range ordered system. The magnetic field is then removed and the thermoremanent magnetization ($M^{tr}$) is measured as a function of temperature. To investigate the TRM for Fe$_{x}$TaS$_2$, we cooled the sample to 3 K from much above $T_c$ under various cooling fields ($H_{cool}$), and in each case, the sample was subsequently heated after the removal of $H_{cool}$ [see Fig.~\ref{TRM}]. We observe that even in zero-field, a significant $M^{tr}$ is present, even for $H_{cool}=$ 100 Oe. We found that as $H_{cool}$ increases, $M^{tr}$ increases, and for $H_{cool} \geq$ 500 Oe, $M^{tr}$ reaches the saturation magnetization, $M_s$. To further confirm TRM in the compound studied, we cooled the sample in negative $H_{cool}$ (-200 Oe and -1 kOe), which provides negative $M^{tr}$ signifying sign reversal under $H_{cool}$. It should be noted that for both negative and positive $H_{cool}$, $M^{tr}$ vanished for $T > 160 K$.   

\par
Considering the irreversible nature of $M$, it is pertinent to study the time evolution of remanent magnetization in the FC state. In this process, the sample was cooled from 200 K under an applied field of $H_{cool}=$ 500 Oe to a suitable temperature ($T_{rlx}$) for relaxation measurement. Subsequently, the field is removed and $M$ was measured as a function of time ($t$). We perform relaxation measurements for several values of $T_{rlx}$, namely 10, 25, 70 K, for an hour. Interestingly, the data at 10 K and 25 K show an insignificant relaxation (around 1\%) over the course of one hour (not shown here). In contrast, the data at 70 K show a large relaxation (more than 10\%), as shown in Fig.~\ref{relax}.

\begin{figure}
	\centering
	\includegraphics[width = 8.5 cm]{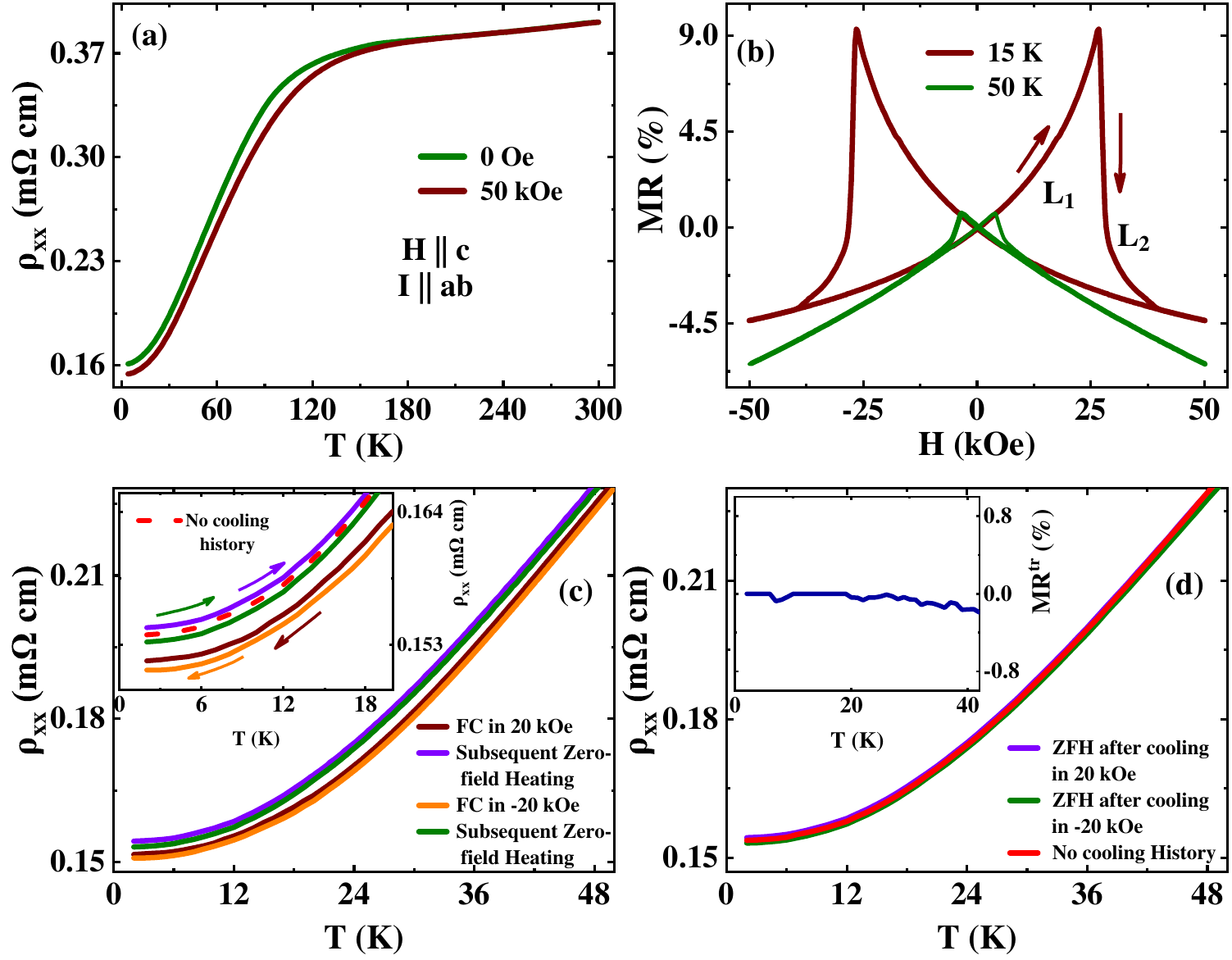}
	\caption{(a) Temperature variation of the longitudinal resistivity under zero applied field (green) and 50 kOe  (wine) are displayed. Both datasets were recorded while cooling the sample from 300 K. (b) shows the field variation of percentage MR at $T =$ 15, 50 and 100 K. (c) Longitudinal resistivity during FC from 300 K to 2 K in 20 kOe and subsequent heating from 300 K to 2 K in zero-field. The same protocol was also repeated in -20 kOe and zero-field. Inset shows the magnified view of this at low $T$, the arrow represents the direction of $T$ change. (d) Here only the zero-field heating curves are plotted, two are from FC procedure in $\pm$ 20 kOe and red line without any FC history. Inset displays the MR obtained after symmetrization due to the TRM at low $T$.}  
	\label{rho_xx}
\end{figure}

\par
To further investigate this, relaxation was measured for different values of $H_{cool}$ at $T_{rlx}=$ 70 K. The data for $H_{cool}=$ 100 and 200 Oe exhibit typical relaxation obeying stretched exponential [$\sim e^{-(t/\tau)^{\beta}}$] behavior. Here, $\tau$ is the time constant for relaxation and $\beta$ is the stretching exponent. However, the normalized relaxations at 500 Oe and 1 kOe show different behavior compared to low-field data. Interestingly, the stretched exponential function cannot fit the full range data for $H_{cool}\geq$ 500 Oe. This deviation can be traced back to the fact that field cooling in $H_{cool}\geq$ 500 Oe produces a fully saturated magnetically arrested state, which is not achieved for $H_{cool} =$ 100 or 200 Oe. However, for $H_{cool} \geq$ 500 Oe, the stretched exponential function can adequately fit the data after the initial 500 seconds of relaxation.

\subsection{Magneto-transport and Hall effect}
The temperature dependence of longitudinal resistivity ($\rho_{xx}$) in zero-field (blue line) and under an applied field of 50 kOe (orange line) for Fe$_{0.27}$TaS$_2$, within the $T$ range of 3 to 300 K, is presented in Fig.~\ref{rho_xx} (a). Measurements were carried out with the current applied in the $ab$ plane and $H$ along the $c$ axis for the 50 kOe data. The data were recorded while cooling from 300 K. In zero field, the resistivity ($\rho_{xx}$) shows an almost linear dependence on $T$ above 150 K, below which it falls rapidly showing typical metallic behavior. Under $H=$ 50 kOe, $\rho_{xx}$ shifts downward, indicating negative magnetoresistance (MR). This behavior is consistent with the reduction of spin-disorder scattering of electrons under an applied magnetic field. 

\par
Furthermore, the field dependence of MR at $T =$ 15 and 50 K is shown in Fig.~\ref{rho_xx} (b) for $H$ applied along the $c$-axis. At $T =$ 15 K, MR increases with $H$. However, with further increases in $H$, MR drops sharply to a negative value within a very narrow range of $H$, close to the coercive fields of the $M$-$H$ and $\rho_{yx}$-$H$ loops [leg L$_2$, see 15 K data of Fig.~\ref{rho_xx} (b)]. Beyond this point, MR decreases linearly as $H$ increases further. When $H$ is reversed, a similar behavior occurs, resulting in a bow-tie-shaped hysteresis loop, as previously reported~\cite{Hardy_PRB,Checkelsky_2008}. Here, leg L$_1$ was obtained after applying $H = -50$ kOe, followed by increasing $H$ toward positive values. Clearly, L$_1$ depicts a positive transverse MR. This occurs because of the bending of the electron trajectory. Eventually, the spins align with the positive field direction at the coercive field (approximately 25 kOe for the 15 K data), causing $\rho_{xx}$ to drop (leg L$_2$) due to coherent scattering of charge carriers and spin moments. As $T$ increases, the width of the bow-tie loop narrows, and the MR evolves toward a more linear variation.

\par
Since Fe$_{x}$TaS$_2$ exhibits a large TRM with $H_{cool}$ as low as 500 Oe, it is important to understand the effect of TRM on MR. To investigate this, $\rho_{xx}$ was measured during the cooling of the sample from 300 K to 2 K under $H_{cool} =$ 20 kOe, as shown by the wine-colored line in Fig.~\ref{rho_xx} (c). After that, the field was removed and $\rho_{xx}$ was measured again (purple line) while heating. Upon removal of $H_{cool}$ at 2 K, $\rho_{xx}$ jumps back to the simple zero-field cooling resistivity curve ($\rho_{xx}^0$), and follows the curve throughout the temperature range. Here, the curve $\rho_{xx}^0$ is obtained by heating the sample in $H=$ 0 after cooling down to 2 K in zero-field.  The same protocol was followed for $H =$ -20 kOe, as shown by the orange and green lines in Fig.~\ref{rho_xx} (c).  A similar behavior is observed for the negative value of $H$.

\begin{figure}
	\centering
	\includegraphics[width = 8.5 cm]{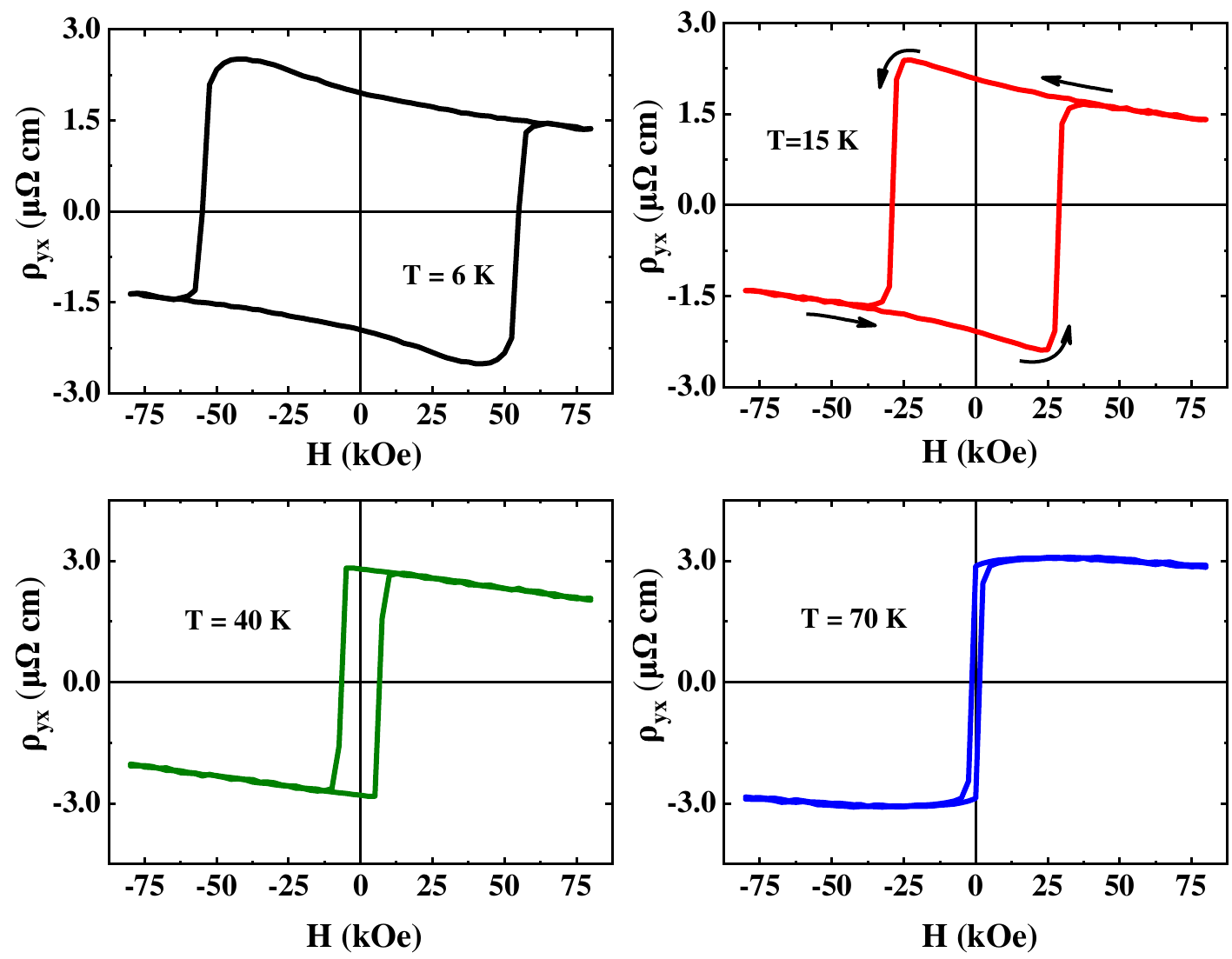}
	\caption{Field variation of the Hall resistivity with $H \parallel c$ measured at different temperatures, after removing the MR contribution through anti-symmetrization. }
	\label{rho_xy}
\end{figure}

\par
To compare the zero-field heating values of $\rho_{xx}$ after FC in $H = \pm$ 20 kOe, we have plotted these values alongside the zero-field $\rho_{xx}$ measured during heating without any cooling history in Fig.~\ref{rho_xx} (d). Surprisingly, these values almost overlap. To further support our claim, we have plotted the thermoremanent magnetoresistance MR$^{tr}$ versus $T$ in the inset of Fig.~\ref{rho_xx} (d), which shows nearly zero values at low $T$, indicating a negligible effect of TRM on $\rho_{xx}$ within the resolution of our experiment. This observation is important for the subsequent sections.

\par
The Hall resistivity ($\rho_{yx}$) of Fe$_{0.27}$TaS$_2$ is measured in the Hall bar geometry by applying $H$ along the $c$ axis and an electric current in the $ab$ plane. The variation of $\rho_{yx}$ with $H$ up to $\pm$ 80 kOe at different temperatures is shown in Fig.~\ref{rho_xy}. In Hall measurement, a slight misalignment of the voltage leads can give rise to unwanted contributions from MR. To eliminate the MR contribution, we have anti-symmetrized $\rho_{yx}$ for positive and negative values of $H$. The $\rho_{yx}$ is found to be highly non-linear and exhibits a large hysteresis loop similar to the magnetization ($M$-$H$) curve, indicating contributions from the AHE. 

\par
For FM materials, $\rho_{yx}$ is generally expressed by the following equation:

\begin{equation}
\rho_{yx} = \rho_{yx}^{OHE} + \rho_{yx}^{AHE}
\end{equation}

where $\rho_{yx}^{OHE}$ is the ordinary Hall effect, which arises due to the lateral motion of electrons due to the Lorentz force in a magnetic field, and $\rho_{yx}^{AHE}$ is the anomalous Hall effect, which has a magnetic origin.

\par
\begin{figure}
	\centering
	\includegraphics[width = 7.5 cm]{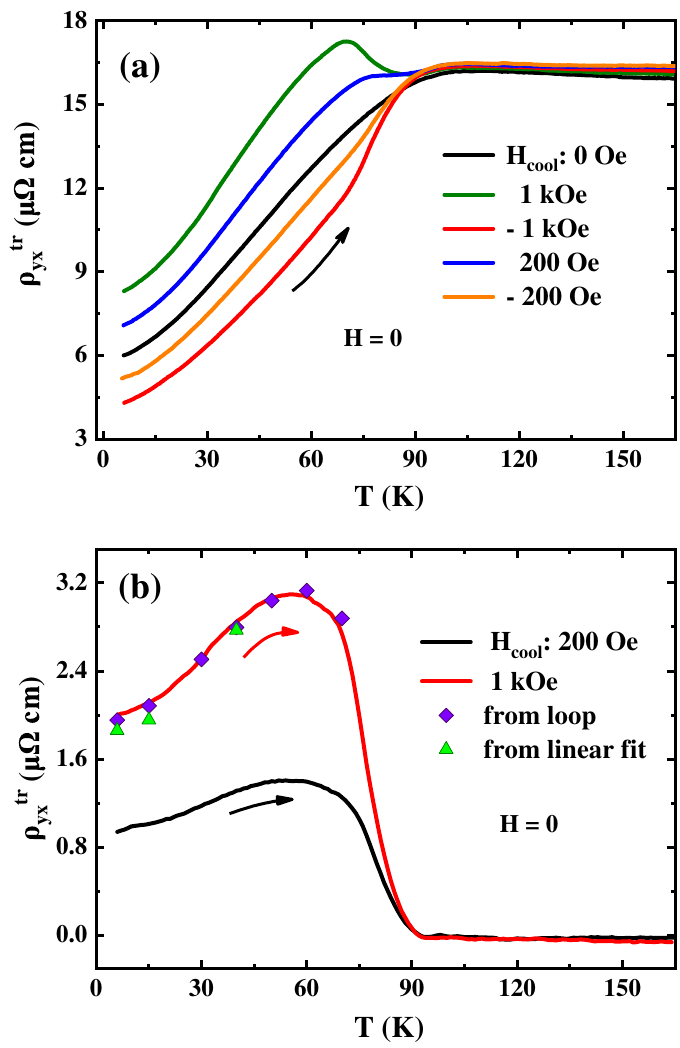}
	\caption{(a) The sample is cooled under different positive and negative fields, referred to as $H_{cool}$. After reaching the lowest temperature (4 K), the field is removed, and $\rho_{yx}$ is measured while heating the sample from 4 K. (b) Represents the variation of Hall resistivity after anti-symmetrization at different $H_{cool}$ values. The symbols (diamond and triangle) show the values of $\rho_{yx}$ obtained from the field variation of $\rho_{yx}$ shown in Fig.~\ref{rho_xy}.}
	\label{spt_rho_xy}
\end{figure}

As shown in Fig.~\ref{rho_xy}, the width of the hysteresis loop in $\rho_{yx}$ decreases as $T$ increases. On the other hand, the saturation value of $\rho_{yx}$ increases with $T$. To estimate $\rho_{yx}^{AHE}$ from the measured $\rho_{yx}$, we employed two different methods. In the first method, we measured the vertical width of $\rho_{yx}$ at a particular temperature at the intercept with the $H = 0$ axis, giving rise to 2$\rho_{yx}^{AHE}$. In the second method, we linearly fitted the high-field saturated part of $\rho_{yx}$ with respect to $T$ and obtained $\rho_{yx}^{AHE}$ from the intercept. The values obtained for $\rho_{yx}^{AHE}$ are shown in Fig.~\ref{spt_rho_xy} (b) using purple diamond and green triangle symbols.

\begin{figure*}
	\centering
	\includegraphics[width =18 cm]{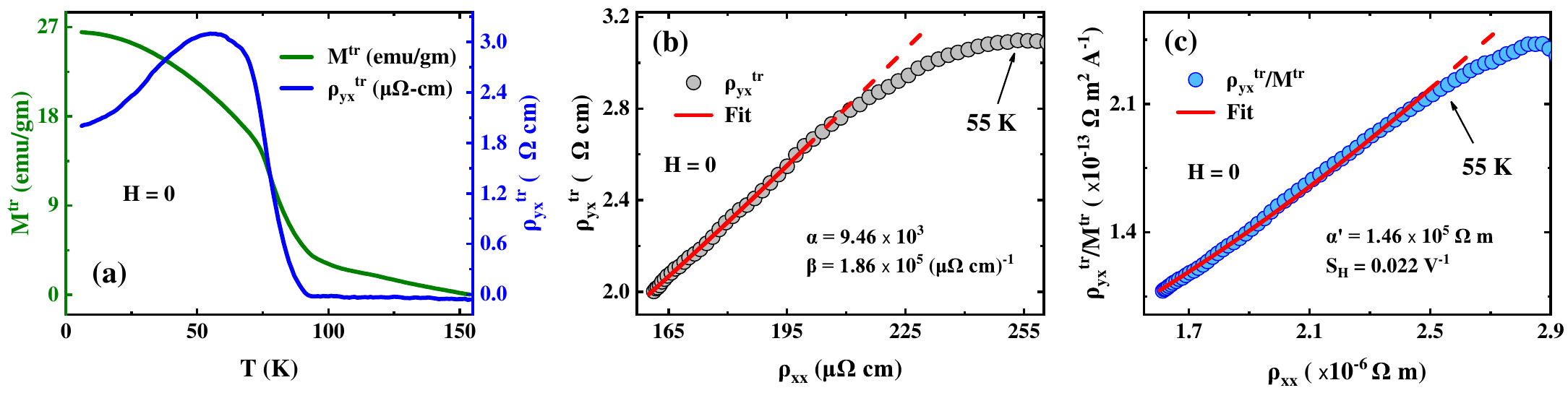}
	\caption{(a) $T$ variation of the magnetization, longitudinal resistivity and Hall resistivity obtained during zero-field heating after a FC with $H_{cool} \geq$ 500 Oe. The superscript `tr' represents thermoremanent as these quantities are originating due to TRM. (b) Plot of $\rho_{yx}^{tr}$ with the variation of $\rho_{xx}$. The red line represents the fitting using eqn.~\ref{eq:rho_scaling1} in the low $T$ region. (c) The variation of $\rho_{yx}^{tr}/M^{tr}$ with longitudinal resistivity without any cooling history ($\rho_{xx}$). The solid red line displays the fit using eqn.~\ref{eq:rho_scaling2}.}
	\label{scaling}
\end{figure*}

\par
The most interesting observation of the study is the effect of TRM in $\rho_{yx}$. To understand this, we have cooled the sample from 180 to 4 K under different applied fields. After reaching 4 K, we set $H$ at zero and measured $\rho_{yx}$ while heating the sample. We call this thermoremanent Hall (TRH) resistivity ($\rho_{yx}^{tr}$).  The $T$ variation of $\rho_{yx}^{tr}$ in different cooling fields is shown in Fig.~\ref{spt_rho_xy} (a). To remove the longitudinal resistivity contribution coming from voltage probe misalignment, we anti-symmetrize the data using the Hall resistivity measured for both positive and negative values of $H_{cool}$, namely, $\rho_{yx} = \frac{1}{2}[\rho_{yx}(+H_{cool}) - \rho_{yx}(-H_{cool})]$~\cite{Hall_symm1,Hall_symm2}. The $T$ variation of the symmetrized $\rho_{yx}^{tr}$ is illustrated in Fig.~\ref{spt_rho_xy} (b). It increases with increasing $T$ and shows a maximum around 57 K then plunges to almost zero above 90 K. Interestingly, the value of $\rho_{yx}^{tr}$ also depends on the cooling field. The black line in Fig.~\ref{spt_rho_xy} (b) is taken, when the applied cooling field was 200 Oe, slightly less to lock the system in the arrested state. On the other hand, the 1 kOe cooling field is sufficient to lock the system in the saturation magnetization state. We compared $\rho_{yx}^{tr}$ with the $\rho_{yx}^{AHE}$ obtained from our $\rho_{yx}$ versus $H$ isotherms [Fig.~\ref{rho_xy}] recorded at various $T$. As depicted in Fig.~\ref{spt_rho_xy} (b), $\rho_{yx}^{tr}$ (shown with continuous line) and $\rho_{yx}^{AHE}$ (as shown by the purple diamond and green triangle symbols) match surprisingly well. This indicates that the Hall resistivity measured in zero-field after suitable field cooling ($H_{cool} \geq$ 500 Oe) is actually the anomalous Hall resistivity, \textit{i.e}, $\rho_{yx}^{tr} \equiv\rho_{xy}^{AHE}$.

\par
To further confirm that the $T$ variation of $\rho_{yx}$ is arising solely from the TRM, we have cooled the sample at a higher field (5 kOe) and measured the $\rho_{yx}$ under a zero-field while heating. The $T$-variation retracts the exact same curve as 1 kOe (not shown here). On the other hand, similar measurement on MR shows no effect in $\rho_{xx}$, i.e. TRM has no effect in $\rho_{xx}$. This establishes that $\rho_{yx}$ is due to TRM in the system. This is a direct proof of the magnetic origin of the AHE present in the system.

\par
To shed more light on the intercorrelation between $M$ with $\rho_{yx}$, we have plotted the $T$ variation of $M$ (green line) and $\rho_{yx}$ (blue line) while heating after a FC with $H_{cool} \geq$ 500 Oe, designated as $M^{tr}$ and $\rho_{yx}^{tr}$ (after anti-symmetrization) respectively in Fig.~\ref{scaling} (a). Since $\rho_{yx}^{tr}$  and $\rho_{yx}^{AHE}$ are equivalent to each other, we can write,
\begin{equation}
    \rho_{yx}^{tr} = \alpha\rho_{xx}+\beta{\rho_{xx}^2}
    \label{eq:rho_scaling1}
\end{equation}
The variation of $\rho_{yx}^{tr}$ with $\rho_{xx}$ is plotted in Fig.~\ref{scaling} (b). The solid red line represents the fit to the data using eqn.~\ref{eq:rho_scaling1}. The fitting model converges for $\alpha = 9.46 \times10^{-3}$ and $\beta = 1.86\times10^{-5}$ $(\mu$$\Omega$-cm$)^{-1}$. This suggests that $\rho_{yx}^{tr}$ has both extrinsic (side-jump and skew scattering) and intrinsic (Berry phase) contributions.

\par
To investigate the effect of TRM on $\rho_{yx}^{tr}$, we have employed the following mathematical expression (see eqn.~\ref{eq:rho_scaling2}), which establishes a connection between the two quantities when no external magnetic field  is applied~\cite{Chapman2013}.

\begin{equation}
    \rho_{yx}^{tr} = S_HM^{tr}{\rho_{xx}^2 \Bigl( 1+\frac{\alpha^{\prime}}{\rho_{xx}} \Bigr)}
    \label{eq:rho_scaling2}
\end{equation}

Here, $S_H$ and $\alpha^{\prime}$ serve as fitting parameters. Fig.~\ref{scaling} (c) presents the plot of $\rho_{yx}^{tr}$ /$M^{tr}$ against $\rho_{xx}$, with the data modeled using eqn. ~\ref{eq:rho_scaling2}. The fitted curve closely matches the experimental trend, revealing a coupled variation between TRH and TRM. The values of the extracted parameter from the fitting are $S_H$ = 0.022 $V^{-1}$ and $\alpha^{\prime} = 1.46\times10^{-6}$ ($\Omega.m$), highlighting the consistency of the model with the observed data.

\begin{figure}
	\centering
	\includegraphics[width =8.5 cm]{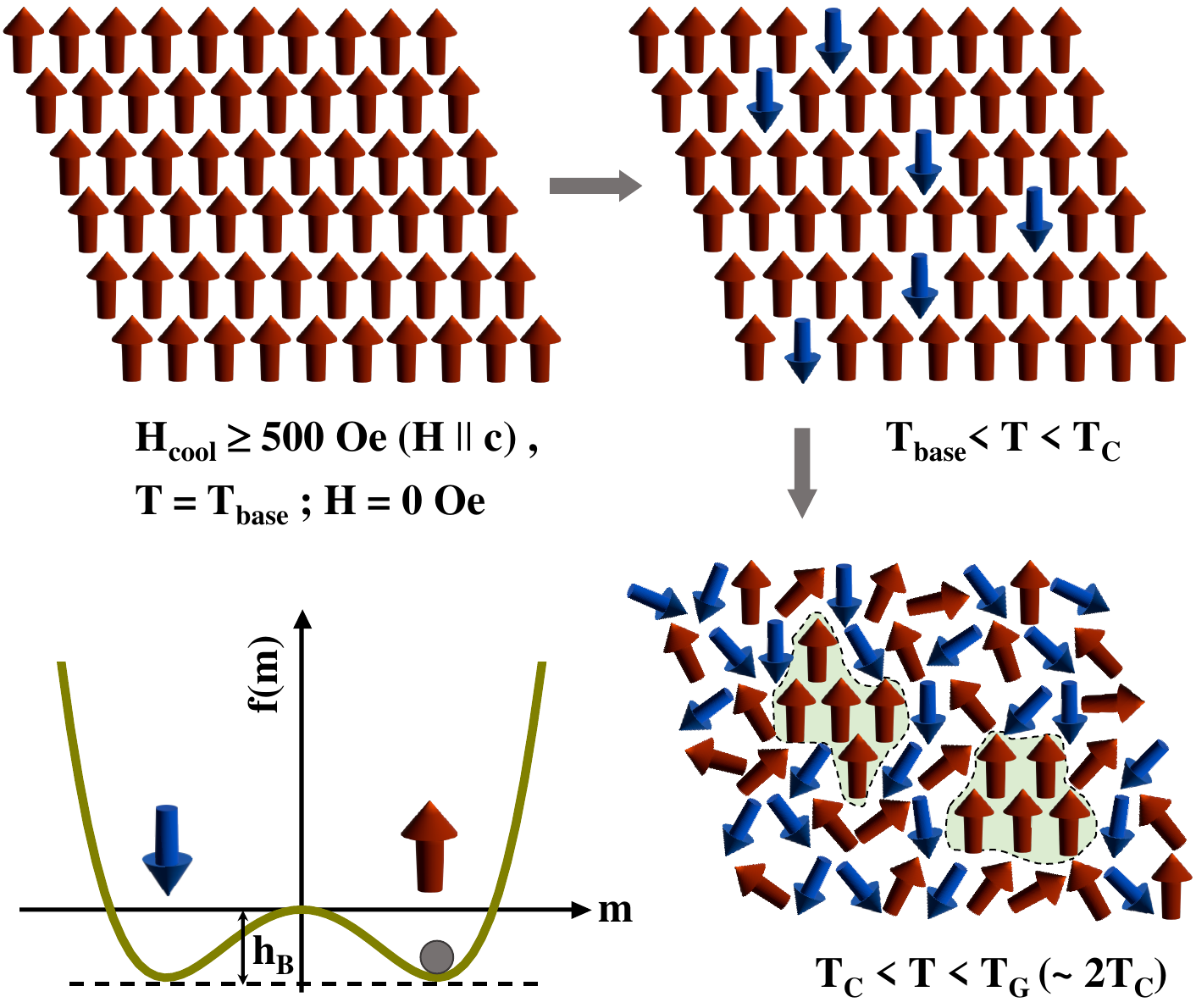}
	\caption{The temperature evolution of the spin configuration after field cooling under an applied field of at least 500 Oe along the $c$-axis. The top-left figure represents the spin configuration at the base temperature ($T_{\text{base}} \approx$ 4 K), where all spins are aligned along the applied field direction. As temperature increases (gray arrow), the system enters an intermediate state ($T_{\text{base}} < T < T_C$), where some spins start to reorient. Above the $T_C$, thermal fluctuations lead to a more disordered spin configuration with Griffiths like FM spin clusters. The bottom-left panel depicts the Landau-type free energy landscape below $T_C$, where two degenerate magnetization states (up and down spins) exist, separated by an energy barrier h$_B$.}
	\label{fig:schematic}
\end{figure}

\section{Discussions} 	 
The structural analysis confirms that Fe$_{x}$TaS$_2$ adopts a hexagonal crystal structure with the \textit{P}6$_3$/mmc space group, closely resembling the structure of its Fe$_{1/4}$TaS$_2$ counterpart~\cite{Morosan2007}. This similarity is strongly supported by SEM-EDX analysis, which indicates an Fe content of approximately 0.27, consistent with the stoichiometry of Fe$_{1/4}$TaS$_2$.
\par
The presently studied Fe$_{0.27}$TaS$_2$ is an Ising ferromagnet that shows a large coercive field and a sharp magnetization reversal. The uniaxial magnetic anisotropy is prevalent in the sample, with the crystallographic $c$ axis being the easy axis of magnetization. Most interestingly, the sample shows a large TRM, when it is cooled under a field greater than 500 Oe. On heating in a zero-field, the TRM diminishes and eventually vanishes above 160 K. The TRM also gives rise to AHE in zero applied field, which is found to be non-zero up to about 90 K.

\par
Such TRM effects are quite common in glassy magnetic systems, where the disordered spin configuration and frustration lead to metastable states that retain the memory of the applied field~\cite{mydosh1993}. However, in the present system, the absence of a glassy magnetic ground state is evident from our field-cooled-field-stop memory measurements (not shown here). This suggests that the TRM effect in this material is driven by a fundamentally different mechanism compared to conventional glassy systems.

\par
There are instances in single crystalline ferro or ferri -magnetic samples without glassy nature, where TRM is observed. For example, TRM is observed in YFeO$_3$, EuS, and FeS$_{1.14}$ (pyrrhotite), and it is associated with a strong magneto-crystalline anisotropy~\cite{Syono1962}. Our Fe$_{x}$TaS$_2$ also belongs to this category, where we observe TRM in the absence of apparent glassiness. 

\par
The mechanism of TRM in our single-crystalline sample can be understood from the nature of the Ising spins. Close to the Curie-point, the magnetic anisotropy is weak, and cooling in a positive field ($H_{cool} \parallel c$) from well above $T_C$ down to 3 K, will produce a single-domain state with all the spin directed along the easy axis. When the field is removed at 3 K, the spins will still remain aligned because the anisotropy energy is high at low temperature (see the spin arrangement in the top left of Fig.~\ref{fig:schematic}). In the Landau-type free energy scenario (see bottom left of Fig.~\ref{fig:schematic}), there exists a free energy barrier $h_B$, between the up spins ($\uparrow$) and the down spins ($\downarrow$). On heating the sample in zero-field with the field-cooling history, the sample does not get demagnetized immediately because of the presence of this free energy barrier. The demagnetization is further restricted, as it is energetically unfavorable to have domains with spins along the hard axis. As we continue to heat the sample, there can be $\uparrow$ to $\downarrow$ switching assisted by thermal energy (see the spin arrangement in the top right of Fig.~\ref{fig:schematic}). Furthermore, the anisotropy energy decreases, enabling the system to have domains in all possible directions, and eventually the TRM  value drops significantly when heated above $T_C$. It is interesting to note that a small but finite TRM still continues to exist even at a temperature as high as 160 K. As discussed in section~\ref{sec:magnetization}, the system shows Griffiths-like phase above in the temperature window $T_C<T<T_G$. Since the studied Fe$_x$TaS$_2$ is an Fe intercalated system, there can be some disorder in Fe occupancy. Additionally, sulfur vacancy is quite common in such chalcogenides. Griffiths instability was originally formulated for the disordered Ising spin system~\cite{Griffiths1969}, and our composition has the necessary attributes for GP. Thus, the observed TRM above $T_C$ is associated with the formation of noninteracting FM clusters in the GP (see the spin arrangement in the bottom right of Fig.~\ref{fig:schematic}).  

\par
The Hall voltage observed in the zero-field heating path after being field cooled is associated with the TRM of the sample. This is evident from the linear nature of the $\rho_{yx}^{tr}/M^{tr}$ vs. $\rho_{xx}$ plot at least below 55 K. We find that $\rho_{yx}^{tr}$ is equally dominated by extrinsic and intrinsic mechanisms.

\par
Although, thermoremanent Hall follows the TRM exceptionally well below $T_C$, a contrasting nature is observed when the sample is heated above $T_C$ = 90 K. TRM continues to have a nonzero value up to a value as high as 160 K due to Griffiths-like short-range correlations, thermoremanent Hall vanishes sharply at $T_C$ [see Fig.~\ref{scaling} (a) for a comparison]. Thus, we observe that TRM is retained in the GP, but it hardly affects $\rho_{yx}^{tr}$. Such an observation is fundamentally important and indicates that a state with broken time-reversal symmetry is essential for the observation of thermoremanent Hall voltage.

\par
In conclusion, large thermoremanence is observed in both $M$ and $\rho_{yx}$ for Fe$_{0.27}$TaS$_2$ when the sample is cooled in a moderate field from $T \sim$ 2$T_C$. The thermoremanent magnetization slowly decreases upon heating in zero-field and eventually vanishes above 160 K. It should be noted that, unlike the Hall effect, no thermoremanence is observed in the magnetoresistance data. The observed thermoremanence can be attributed to the strong magnetocrystalline anisotropy present in the sample. In particular, although TRM persists above $T_C$, $\rho_{yx}^{tr}$ becomes zero above this temperature. This highlights that while the thermoremanence survives into the Griffiths phase, it does not affect the anomalous Hall resistivity. Therefore, it is evident that the non-interacting ferromagnetic clusters in the Griffiths phase are crucial for TRM, whereas $\rho_{yx}^{tr}$ remains insensitive to these clusters above $T_C$.

\section*{Acknowledgments}
M.N. would like to thank CSIR, India, for his research fellowship [File No. 09/080(1131)/2019-EMR-I]. The UGC-DAE CSR, Kolkata centre is acknowledged for providing the necessary facilities. The authors gratefully acknowledge access to the X-ray facilities of the Materials Characterization Laboratory at the ISIS Facility.

\bibliography{ref_FTS}
\bibliographystyle{apsrev4-2}
	
\end{document}